\documentclass[a4paper]{article}
\usepackage{a4wide}
\usepackage[utf8]{inputenc}
\usepackage[T1]{fontenc}
\usepackage[english]{babel}
\usepackage{amsmath}
\usepackage{graphicx}
\usepackage{subcaption}
\usepackage{textcomp}
\usepackage{cite}
\usepackage{url}
\usepackage{wrapfig}
\usepackage{amssymb}
\usepackage{ulem}
\usepackage{epstopdf}
\usepackage{epsfig}
\usepackage{color}

\begin{document}
\title{Niobium stripline resonators for microwave studies on superconductors}

\author{Markus Thiemann$^1$, Daniel Bothner$^2$, Dieter Koelle$^2$, \\ Reinhold Kleiner$^2$, Martin Dressel$^1$, Marc Scheffler$^1$ }
\date{}
\maketitle

$^1$1.Physikalisches Institut, Universität Stuttgart, Pfaffenwaldring 57, D-70550 Stuttgart, Germany\\
$^2$ Physikalisches Institut and Center for Collective Quantum Phenomena in LISA$^+$, Universität Tübingen, Auf der Morgenstelle 14, D-72076 Tübingen, Germany

%ad{scheffl@pi1.physik.uni-stuttgart.de}

\begin{abstract}
%Superconducting stripline microwave resonators, where the material under study constitutes one of the ground planes, are a highly sensitive experimental tool to investigate the intrinsic properties of superconducting bulk samples. 
Microwave spectroscopy is a powerful experimental tool to reveal information on the intrinsic properties of superconductors. Superconducting stripline  resonators, where the material under study constitutes one of the ground planes, offer a high sensitivity to investigate superconducting bulk samples. 
In order to improve this measurement technique, we have studied stripline resonators made of niobium, and we compare the results to lead stripline resonators. With this technique we are able to determine the temperature dependence of the complex conductivity of niobium and the energy gap $\Delta(0)=2.1$ meV. Finally we show measurements at the superconducting transition of a tantalum bulk sample using niobium stripline resonators.   
\end{abstract}

\section{Introduction}
Optical spectroscopy on metals and superconductors can reveal information about charge carrier dynamics as well as electromagnetic excitations \cite{Dressel2002a}. In the case of superconductors and correlated metals, these are connected to low energy scales both in temperature and in frequency and therefore have to be probed with low-energy optics, i.e.\ in the THz and GHz ranges \cite{Scheffler2013}. Obtaining spectral information on highly conductive materials at cryogenic temperatures in the microwave range has proven technically challenging. While broadband measurements on a number of superconductors \cite{Booth1995,Ohashi2006,Steinberg2008,Pompeo2010,Liu2011} have successfully been performed on thin films using Corbino spectrometers \cite{Booth1994,Scheffler2005a,Steinberg2012} or on single crystals using a bolometric approach (without obtaining phase information) \cite{Turner2003,Turner2004,Bobowski2010}, there is so far no experimental technique that allows broadband, phase-sensitive measurements on low-loss bulk samples. Therefore, in recent years new techniques have been developed that are resonant in nature, thus are very sensitive to small losses in bulk samples, and at least give certain information about the frequency-dependent response by operating at several resonant frequencies \cite{Huttema2006,Ormeno2006,Scheffler2013}. Our approach is based on superconducting striplines where the sample of interest replaces one of the ground planes \cite{DiIorio1988}. So far, our striplines were made of Pb thin films \cite{Scheffler2012,Hafner2014}. In the present work, we investigate in which regard Nb thin films can also be employed for this purpose and whether their performance can even surpass that of the Pb films.

\section{Experiment}
The stripline is a layered structure, consisting of the center strip sandwiched between two dielectric plates and two conducting ground planes (see Fig. 1(a)).
\begin{figure}[h]
\begin{subfigure}[b]{0.3\linewidth}{\includegraphics[width=\linewidth]{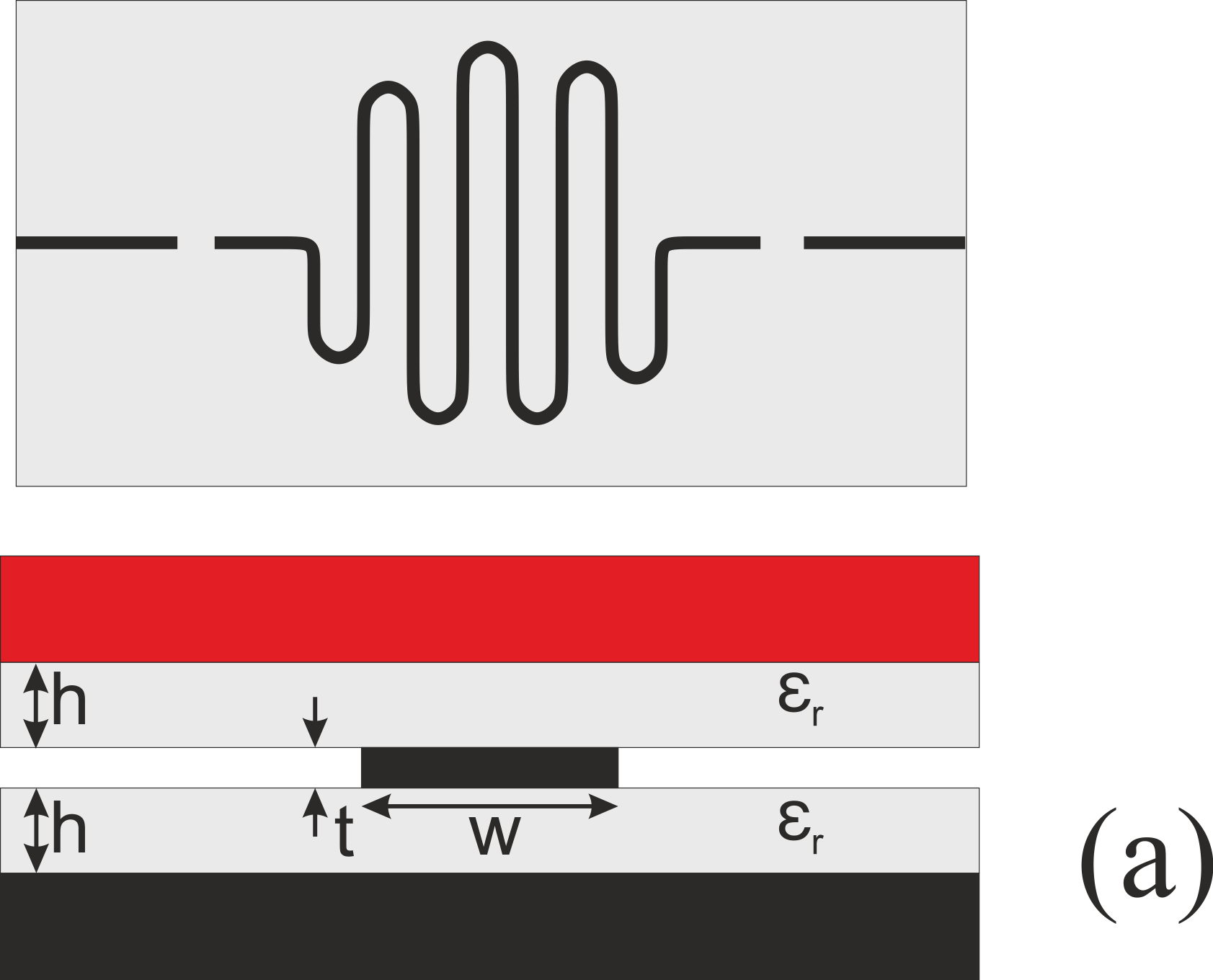}}
%\caption{}
\label{crosssection}
\end{subfigure} \hspace{0.2\linewidth}
\begin{subfigure}[b]{0.4\linewidth}{\includegraphics[width=\linewidth]{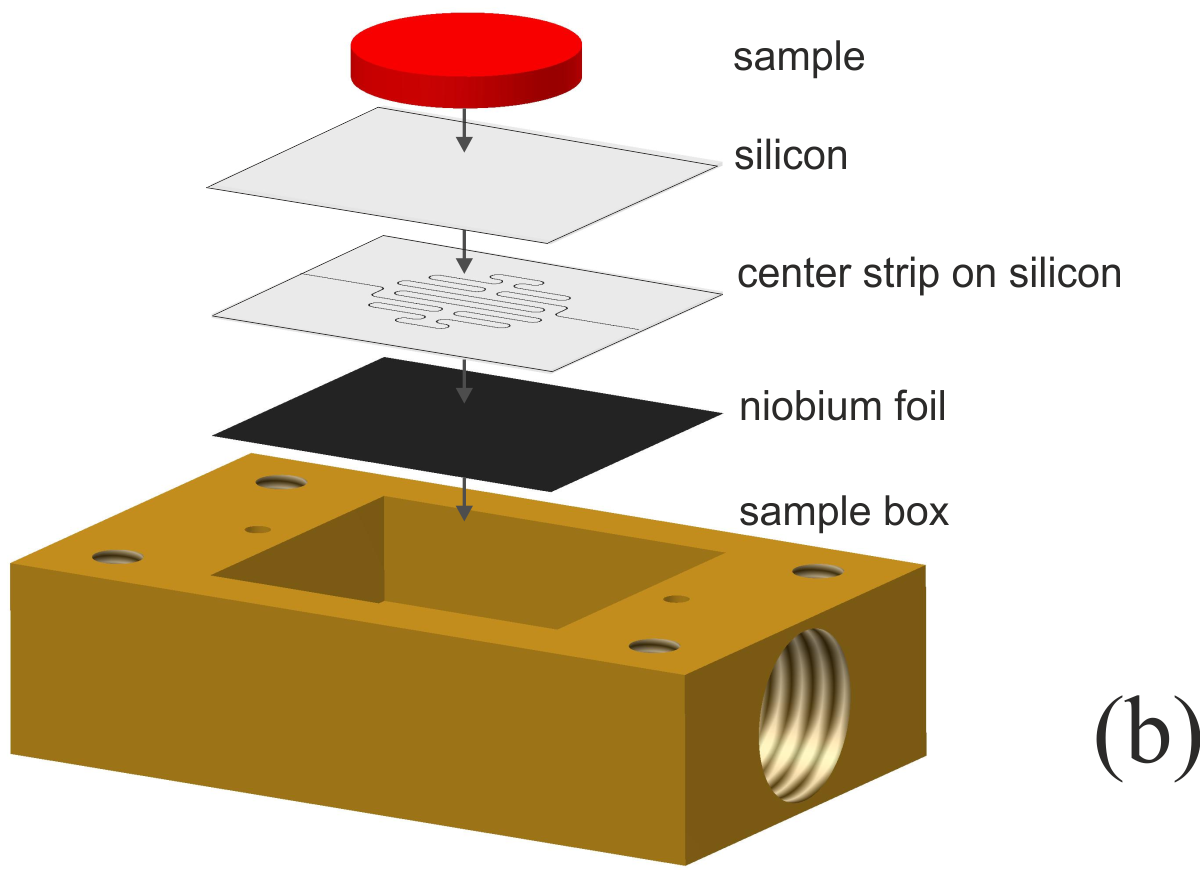}}
%\caption{}
\label{assembly}
\end{subfigure}
\caption{(a) Top: Top view on the meandered center strip with the gaps. Bottom: Cross section of a stripline. The characteristic impedance of a stripline is a function of height $h$ of the dielectric planes, width $w$ and height $t$ of the center strip and the dielectric constant $\epsilon_r$ of the dielectric \cite{Wheeler1978}. (b) Assembly of the stripline resonator into the brass sample box.}
\end{figure}
In case of the Nb stripline resonators, two 100 $\mu$m thin 12x10 mm$^2$ silicon plates were used as a dielectric, with a resistivity >5000 $\Omega$/cm, due to the low dielectric losses and high dielectric constant \cite{Krupka2006} and 8 $\mu$m thick Nb foils as ground planes. Nb ($T_c$=9.2 K) was used as the conductor material instead of the previously used Pb ($T_c$=7.2 K) due to the higher transition temperature. The center strip was fabricated by sputtering Nb onto one of the silicon plates, followed by optical lithography and SF$_6$ etching. Due to two 80 $\mu$m gaps in the center conductor the stripline becomes a resonant structure. Since the frequency of the fundamental mode depends on the length of the stripline section between the gaps, the center strip was meander-shaped as shown in Fig. 1(a) to achieve resonance frequencies as low as possible for a given sample size.
The stripline resonator was mounted inside a brass box and connected via coaxial microwave connectors to the 50 $\Omega$ measurement circuitry (see Fig. 1(b),\cite{Hafner2014}). By replacing one ground plane by any conducting sample, it is possible to determine the frequency and temperature dependence of the surface resistance $R_s$ and the penetration depth $\lambda$ of the sample \cite{Hafner2014}.
In the present study all-Nb resonators and Nb resonators loaded with a tantalum sample were investigated in a temperature range between 1.6 K and 9.2 K.

\section{Results}
\subsection{All-Nb resonator}
In Fig. 2(a) the frequency dependence of $R_s$ (obtained for 6 resonances at frequencies between 1.5 GHz and 11 GHz) of an all-Nb resonator is compared to $R_s$ of an all-Pb resonator. Both show a quadratic frequency dependence in accord with theoretical predictions \cite{Halbritter1974}. Surprisingly $R_s$ of the Nb remains in the same order of magnitude as that of the Pb, despite the higher $T_c$. 
It has already been shown that defects on the resonator surface can increase $R_s$ \cite{Pierce1973}. Therefore microscopic non-superconducting defects in the sputtered center strip may be an explanation for the higher $R_s$ \cite{Benvenuti2001}.
\begin{figure}[h]
\begin{subfigure}{0.5\linewidth}{\includegraphics[width=\linewidth]{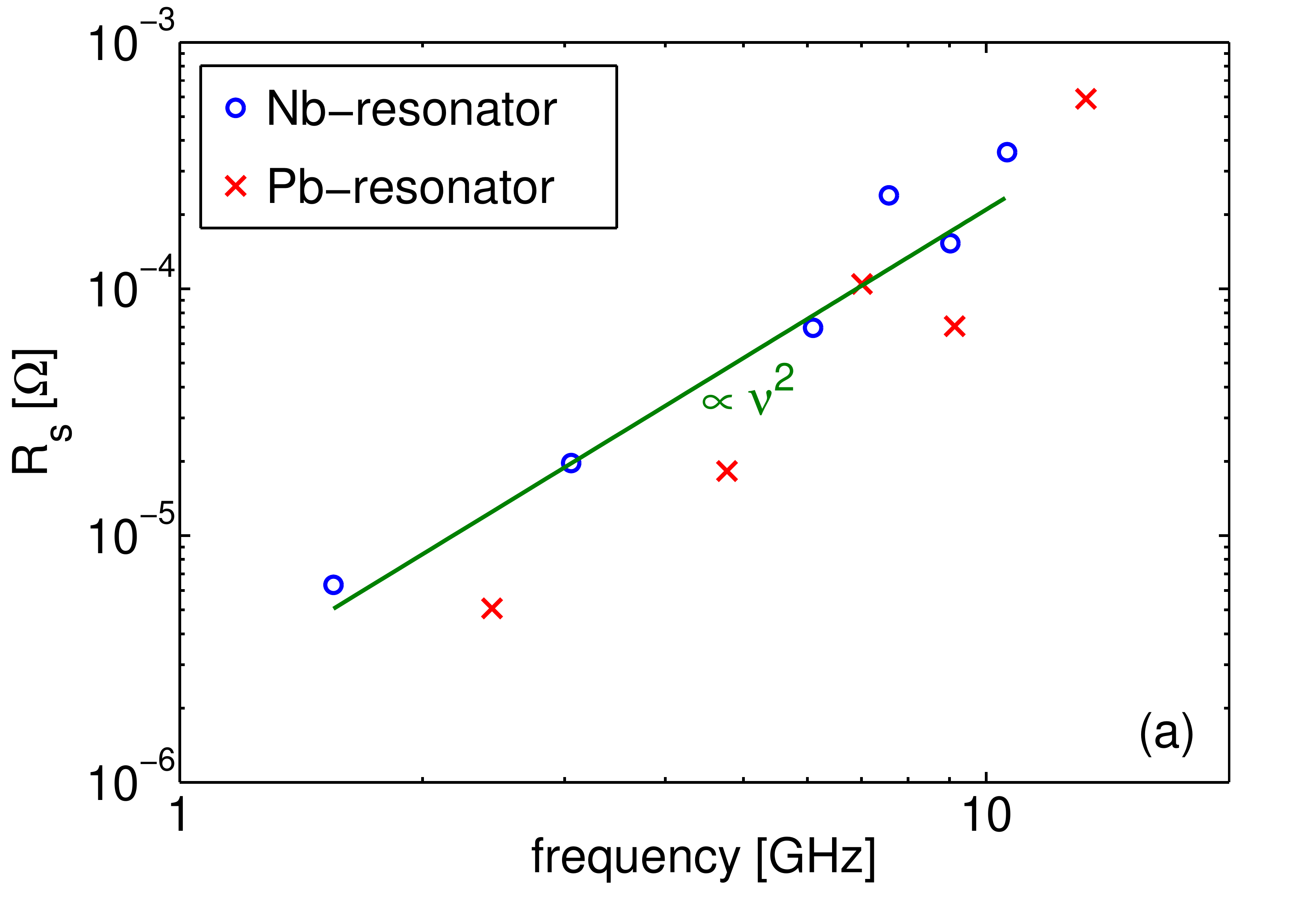}}
%\caption{}
\label{comp_Pb_Nb}
\end{subfigure}
\begin{subfigure}{0.5\linewidth}{\includegraphics[width=\linewidth]{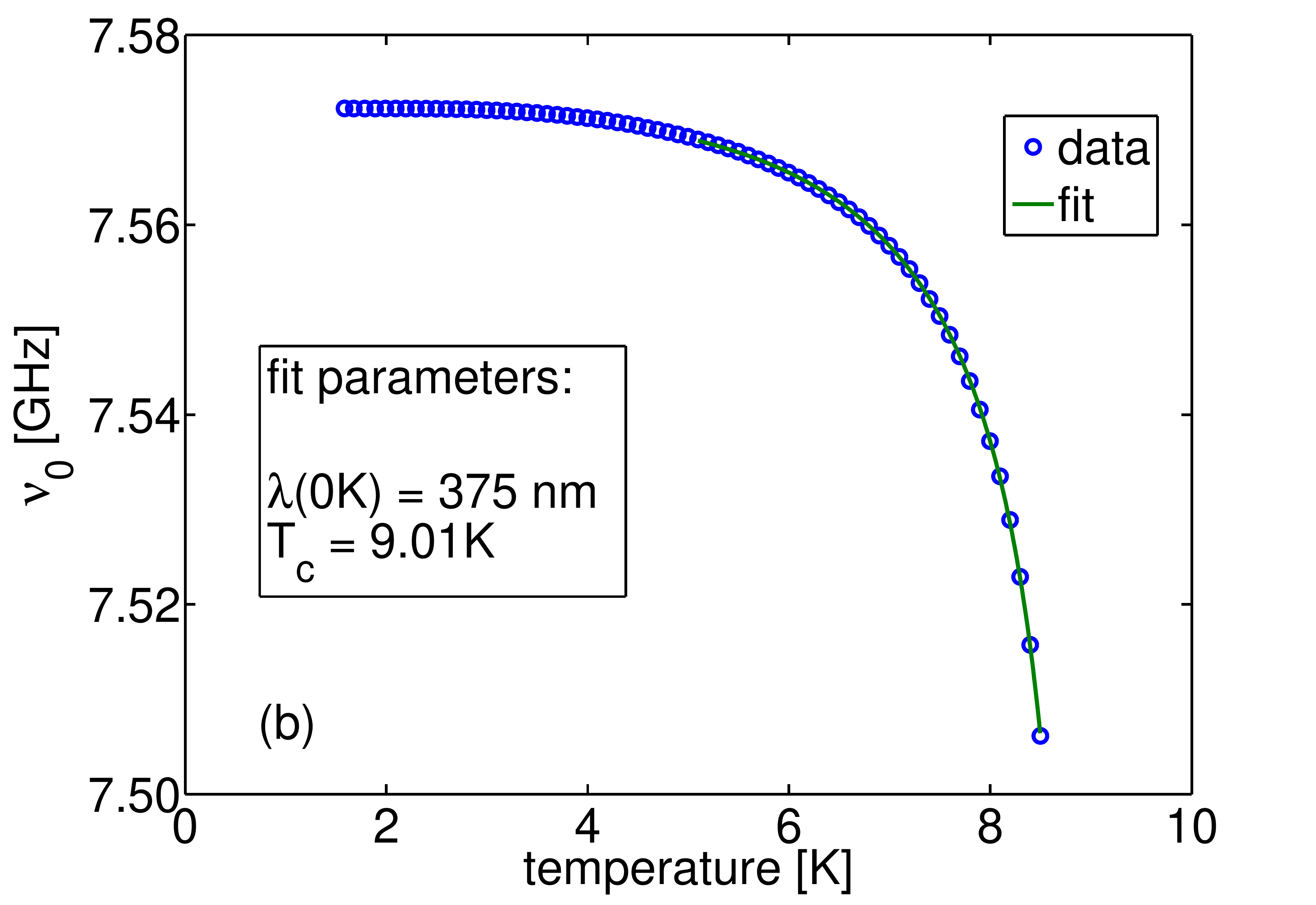}}
%\caption{}
\label{Nb_freq_temp}
\end{subfigure}
\caption{(a) Frequency dependence of $R_s$ of Nb and Pb at $T=1.6$ K. Despite the higher $T_c$, Nb shows a higher $R_s$. (b) Temperature dependence of the resonance frequency of the 5$^\mathrm{th}$ mode. The green line represents a fit from which $\lambda(0\mathrm{K})$ and $T_c$ can be determined as described in \cite{Hafner2014}.}
\end{figure}
The measured temperature dependence of the resonance frequency is shown in Fig. 2(b). The drop of $\nu_0$ towards higher temperatures is due to the Nb entering the normal conducting state. The temperature dependence of $\nu_0$ gives access to the penetration depth $\lambda (0 \, \textrm{K}) \approx 377$ nm and from that the imaginary part of the impedance $X_s$ can be determined \cite{Dressel2002a}. In Fig. 3(a) the temperature dependence of the impedance is shown.
\begin{figure}[h]
\begin{subfigure}{0.5\linewidth}{\includegraphics[width=\linewidth]{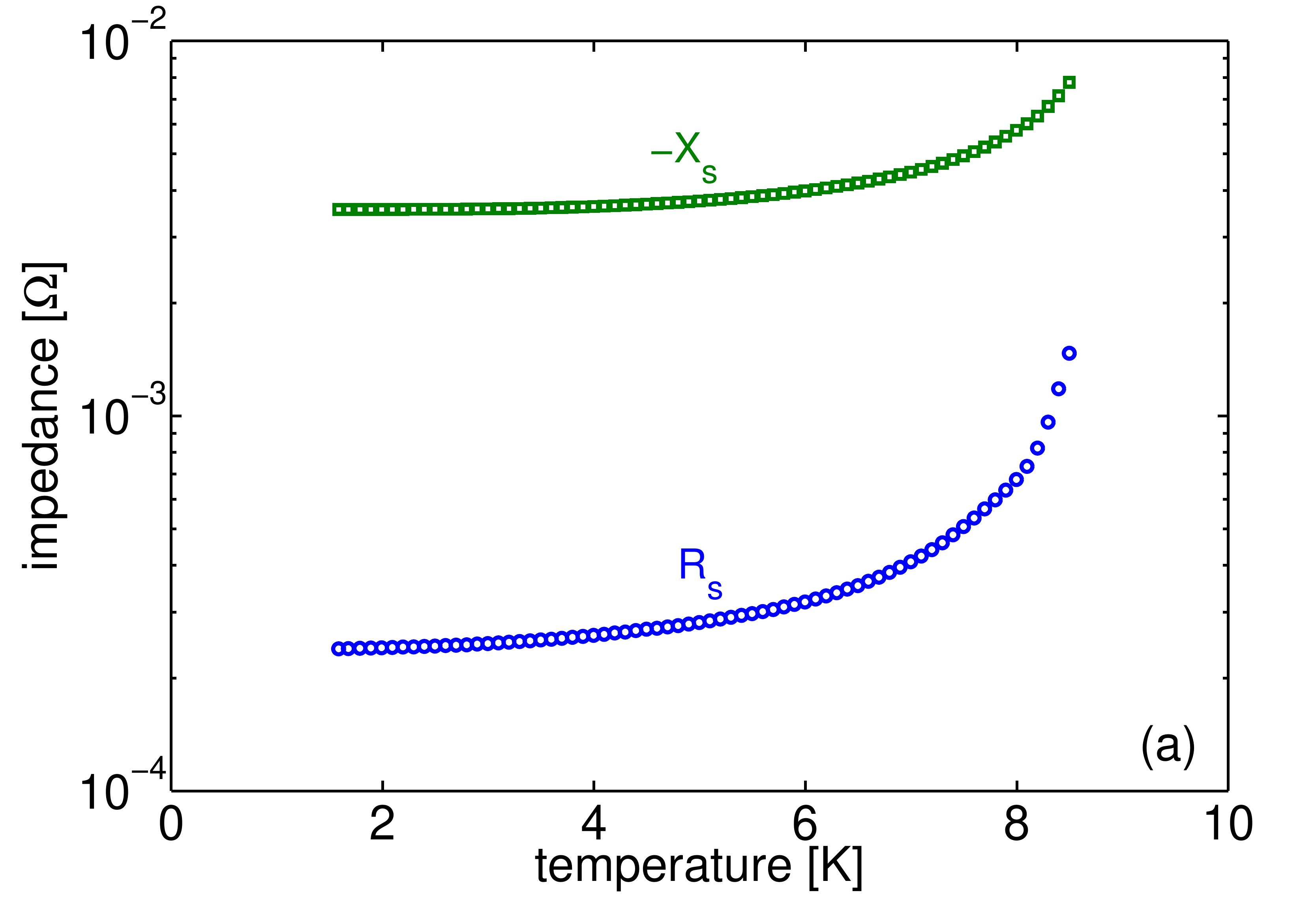}}
%\caption{}
\label{Z_T}
\end{subfigure}
\begin{subfigure}{0.5\linewidth}{\includegraphics[width=\linewidth]{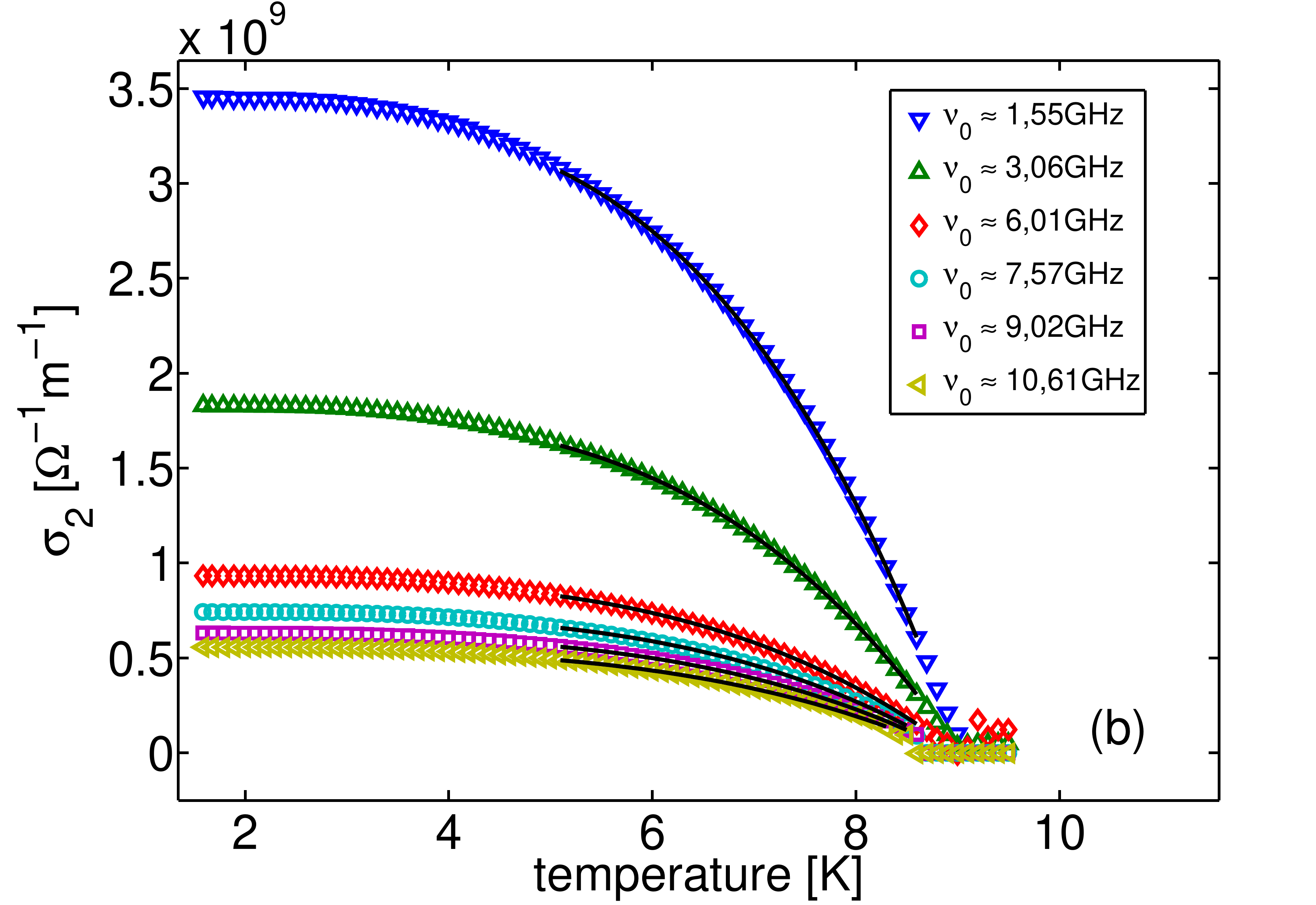}}
%\caption{}
\end{subfigure}
\caption{(a) Temperature dependence of the real and imaginary parts of the complex impedance of a pure Nb resonator obtained from the 5$^{th}$ mode. (b) Temperature dependence of the imaginary part $\sigma_2$ of the complex conductivity obtained from the fundamental mode and higher harmonics. The black lines represent fits of Eq.~(\ref{energygap}).}
\label{sigma2}
\end{figure}
The complex conductivity $\hat{\sigma}=\sigma_1+i\sigma_2$ can now be calculated from the complex impedance \cite{Dressel2002a}. Fig. 3(b) shows  the temperature dependence of the imaginary part $\sigma_2$ of the conductivity, which is connected to the superconducting energy gap $\Delta$ via:
\begin{equation}
\frac{\sigma_2(T)}{\sigma_n} \approx \frac{\pi \Delta(T)}{\hbar \omega}\tanh\left(\frac{\Delta(T)}{2k_\mathrm{B}T}\right)
\label{energygap}
\end{equation}
By using the high temperature approximation $\Delta(T) \approx \Delta(0\mathrm{K}) \sqrt{3.016\left(1-\frac{T}{T_c}\right)-2.4\left(1-\frac{T}{T_c}\right)^2}$ \cite{Ferrell1964}, Eq.~(\ref{energygap}) can be fitted to the measured data with the fit parameters $\Delta(0\mathrm{K})$, $T_c$, and the real part $\sigma_n$ of the conductivity in the normal conducting state. Averaging $\Delta(0\mathrm{K})$ obtained from all measured modes gives a value of 2.1 meV or $2\Delta(0\mathrm{K})=5.3 k_\mathrm{B}T_c$. The energy gap of Nb thin films has already been investigated with THz-spectroscopy and a value of $2\Delta(0)=4.1k_\mathrm{B}T_c$ was observed \cite{Pronin1998}. Both studies show deviations from the weak coupling BCS limit ($2\Delta(0\mathrm{K})=3.5k_\mathrm{B}T_c$ \cite{Kleiner}) and may indicate a strong electron-phonon interaction.
\subsection{Nb resonator loaded with tantalum sample}
As a further test measurement we present data obtained from a bulk tantalum sample. Tantalum has a suitable $T_c=4.5$ K which lies between the transition temperature of the Nb resonators and the lowest temperature accessible with our $^4$He cryostat. 
\begin{figure}[h]
\begin{subfigure}{0.45\linewidth}{\includegraphics[width=\linewidth]{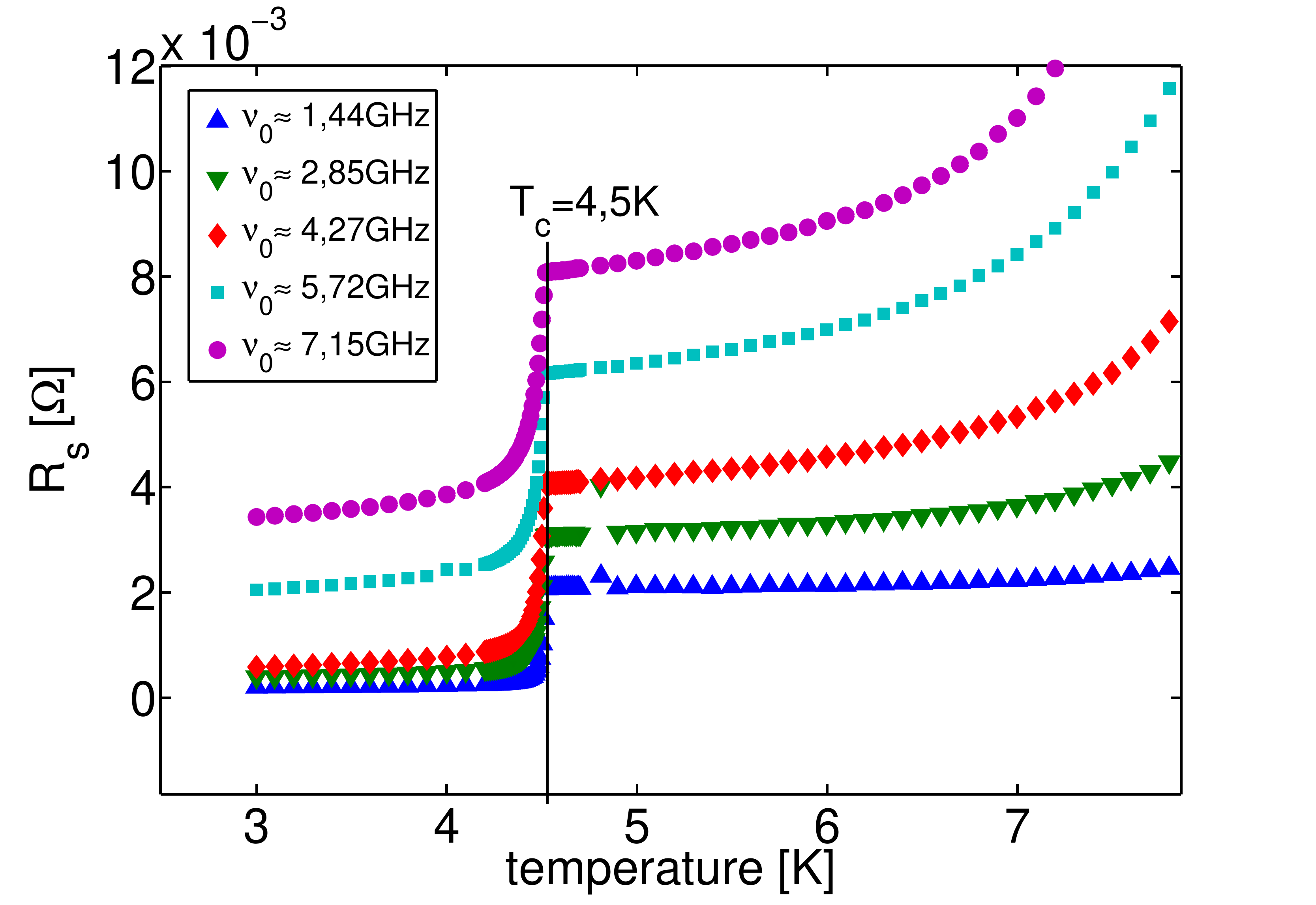}}
%\caption{}
\label{Ta_RsT}
\end{subfigure}
\begin{subfigure}{0.45\linewidth}{\includegraphics[width=\linewidth]{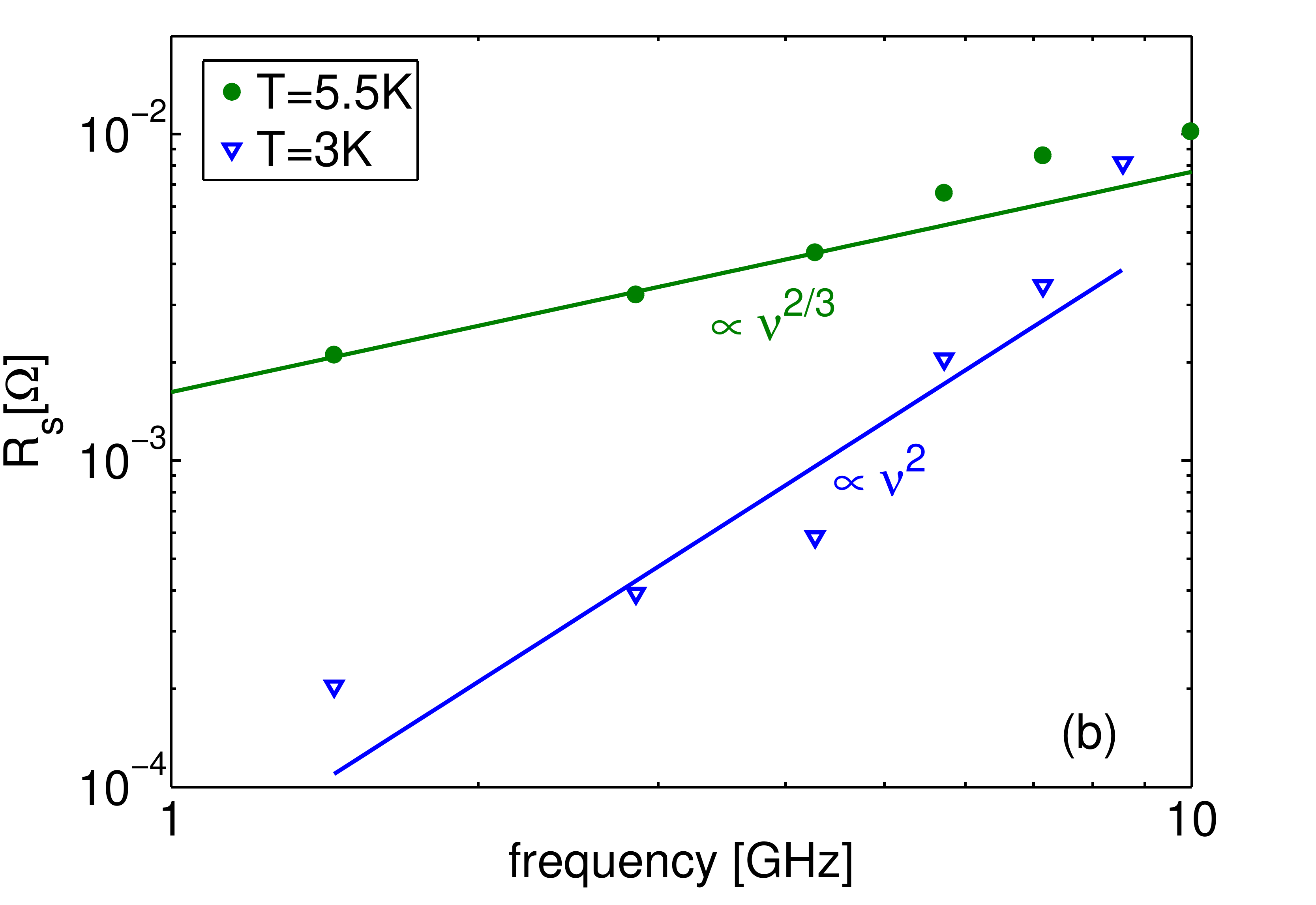}}
%\caption{}
\label{Ta_sup-normal}
\end{subfigure}
\caption{(a) Temperature dependence of $R_s$ of tantalum for several modes. The drop in $R_s$ at 4.5 K is due to the tantalum entering the superconducting state. (b) Frequency dependence of $R_s$ of tantalum in the metallic and superconducting states. Lines are guides to the eye, representing $\nu^{2/3}$ and $\nu^2$ dependencies.}
\end{figure}
Fig. 4(a) shows the temperature dependence of $R_s$ of the tantalum sample obtained from the different modes. Clearly a drop in $R_s$ is visible at 4.5 K, which is due to the tantalum entering the superconducting state. At a temperature of 5.5 K tantalum is in the metallic state and $R_s$ shows a frequency dependence of $R_s(\nu) \propto \nu^{2/3}$, which indicates that the tantalum is in the anomalous skin effect regime (see Fig. 4(b)) \cite{Hafner2014}. At 3 K the tantalum sample has entered the superconducting state, and a clear change in the frequency dependence is visible. 
%\begin{figure}[h]
%\begin{subfigure}{0.5\linewidth}{\includegraphics[width=\linewidth]{pictures/mode3f0vsT.eps}}
%\caption{}
%\label{Ta_f0T}
%\end{subfigure}
%\begin{subfigure}{0.5\linewidth}{\includegraphics[width=\linewidth]{pictures/Zr_f0vsT.eps}}
%\caption{}
%\label{Zr_f0T}
%\end{subfigure}
%\caption{(a)Temperature dependence of $\nu_0$ of a Nb-resonator loaded with a tantalum sample. The drop in $\nu_0$ towards higher temperatures is due to the niobium entering the metallic state. The inset shows a magnification of the red boxed section. At the $T_c=4.5$ K a dip in the resonance frequency is visible. (b)Temperature dependence of $\nu_0$ of a lead resonator loaded with a  Zr sample measured in a dilution refrigerator. Also here a dip is visible in $\nu_0$ at the $T_c$ of the Zr sample.}
%\end{figure}

\section{Conclusions and outlook}
Using Nb as a superconducting material for stripline resonators, we increased the upper limit of the measurable temperature range from 6 K to 8 K, but a lower $R_s$ was not achieved. Future optimization in film growth might reduce the surface impedance and then allow even more sensitive measurements on low-loss bulk superconductors. In contrast to the previous studies using Pb resonators \cite{Hafner2014}, Nb is a type-II superconductor that might allow operation in higher magnetic fields. This might be aggravated by field-dependent microwave losses due to vortices, but strategies to overcome these problems are being developed \cite{Bothner2011, Bothner2012}. The test measurements on superconducting tantalum prove the applicability of our technique. Since our approach can easily be implemented in a dilution refrigerator, microwave studies on a number of exotic superconductors, e.g. heavy-fermion superconductors \cite{Scheffler2013,Truncik2013} or Sr$_2$RuO$_4$ \cite{Ormeno2006} now become feasible.    
\section*{Acknowledgements}
We thank Gabriele Untereiner and Conrad Clau\ss{} for experimental support, and we acknowledge financial support by the DFG, including SFB/TRR21.

\end{document}